\documentclass[11pt]{article}
\usepackage{vietnam,epsfig}

\def\be{\begin{equation}}
\def\ee{\end{equation}}
\def\bea{\begin{eqnarray}}
\def\eea{\end{eqnarray}}

\begin{document}
\vspace*{4cm}
\title{AN UPPER LIMIT TO PHOTONS FROM FIRST DATA
TAKEN BY THE PIERRE AUGER OBSERVATORY}

\author{Markus Risse \footnote{Now at: 
University of Wuppertal, Department of Physics,
42097 Wuppertal, Germany;
\\ electronic address: risse@physik.uni-wuppertal.de
} for the Pierre Auger Collaboration \footnote{The full author list
is available at http://www.auger.org/admin .}}

\address{Forschungszentrum Karlsruhe, Institut f\"ur Kernphysik,
76021 Karlsruhe, Germany}

\maketitle\abstracts{
Many models for ultra-high energy cosmic rays postulate exotic 
scenarios to explain the sources or the nature of these
particles. A characteristic feature of these models is the prediction
of a significant flux of photons at ultra-high energy. The Pierre Auger
Observatory offers a great potential to search for such photons.
We present shower observables with sensitivity to photons and the
search strategy employed. An upper limit to photon primaries is derived
from first Auger data. Prospects for constraining theoretical source
models are discussed.}

\section{Introduction}

The origin of ultra-high energy (UHE) cosmic rays above $10^{19}$~eV
(= 10 EeV) is still unknown.\cite{reviews}
A ``smoking gun'' of non-acceleration (top-down) models of UHE cosmic-ray
origin is the observation of UHE photons.
These models (Super Heavy Dark Matter,\cite{shdm} 
Topological Defects,\cite{td} Z-Bursts,\cite{zb} ...)
were invoked in particular to account for a possible 
continuation \cite{agasa-gzk} of the
cosmic-ray flux above $E_{\rm GZK} \sim 6 \times 10^{19}$~eV
without the flux suppression expected from photo-pionproduction of nucleons
on the microwave background.\cite{gzk}
From considerations of QCD fragmentation,\cite{frag} copious UHE photons
are predicted to be generated as secondaries in the decay or annihilation
chains of the proposed particles and interactions.
Photon fractions in the cosmic-ray flux at the Earth of $\sim$10\% above
10~EeV and $\sim$50\% above 100~EeV would 
result.\cite{bhat-sigl,sarkar03,models,ellis}
Based on first data registered by the Pierre Auger Observatory,\cite{auger}
we present an upper limit to the fraction of UHE photons by comparing
the observed air showers to calculations assuming photons as primaries.
Further details of this analysis can also be found in Ref. \cite{paper}.

\begin{figure}
\begin{center}
\epsfig{figure=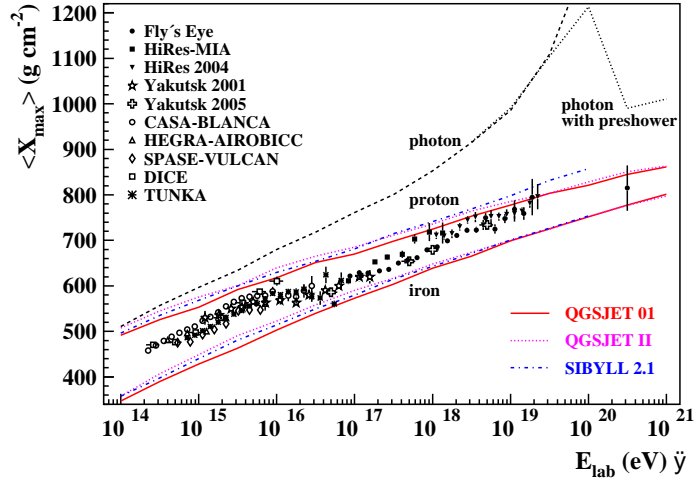,height=2.5in}
\caption{
Average depth of shower maximum $<$$X_{\rm max}$$>$ versus energy
simulated for primary photons, protons and iron nuclei.
Depending on the specific particle trajectory through the geomagnetic
field, photons above $\sim  5 \times 10^{19}$~eV can create a
preshower: as indicated by the splitting of the
photon line, the average $X_{\rm max}$ values then do not only depend
on primary energy but also on arrival direction.
For nuclear
primaries, calculations for different hadronic interaction models
are displayed (QGSJET 01,$^{17}$ QGSJET II,$^{18}$
SIBYLL 2.1 $^{19}$).
Also shown are experimental data (for references to the experiments,
see Ref. $^{20}$).
\label{fig-xmaxvse}}
\end{center}
\end{figure}

Giant air shower experiments are well suited to search for UHE photons.
Similar to nuclear primaries and unlike neutrino primaries, the atmospheric
overburden is large enough for UHE photons to initiate a well-observable
particle cascade. 
Certain observables in photon-induced showers are expected to show
distinct differences compared to those in nuclear primary showers.
This is firstly, because for nuclear primaries it takes several
cascading steps until most energy is transferred to electrons and photons,
while photons initiate an almost purely electromagnetic cascade.
Secondly and connected to this, the high-energy processes of LPM~\cite{lpm}
and preshower~\cite{erber,homola} effect
strongly modify photon showers but not nuclear ones at 10-100~EeV.
In particular, UHE photon showers are expected to reach their maximum
at significantly larger depths, see Figure~\ref{fig-xmaxvse}.
In this work, we obtain a photon limit from the {\em direct} observation
of the shower profile with fluorescence telescopes,
using the depth of shower maximum $X_{\rm max}$ as the
discriminating observable.

\section{Data}

The Auger data used in this analysis were taken with a total
of 12 fluorescence telescopes situated at
two different sites,\cite{bellido} during the period January 2004
to February 2006. The number of surface detector
stations deployed \cite{bertou05} grew during this period
from about 150 to 950.
To achieve a high accuracy in reconstructing the shower geometry,
we make use of the ``hybrid'' detection technique, i.e. we
select events observed by both the ground array and the
fluorescence telescopes.\cite{mostafa}
A detailed description of the Auger Observatory is given
in.\cite{auger}

Cascading of photons in the geo\-magnetic field \cite{erber}
is simulated with the PRE\-SHO\-WER code~\cite{homola} and
shower development in air, including the LPM effect,\cite{lpm} is
calculated with CORSIKA.\cite{heck}
For photo-nuclear processes, we assume the extrapolation of
the cross-section as given by the Particle Data Group,\cite{pdg} and we employed
QGSJET~01 \cite{qgs01} as a hadron event generator.

The reconstruction of the shower profiles \cite{bellido,argiro}
is based on an end-to-end calibration of the
fluorescence telescopes.\cite{brack}
Monthly models for the atmospheric density profiles are used which
were derived from local radio soundings.\cite{keilhauer}
An average aerosol model is adopted based on measurements of the
local atmospheric aerosol content.\cite{roberts}
Cloud information is provided by IR monitors, positioned at
the telescope stations.\cite{roberts}
Cross-checks on clouds are obtained from
measurements with LIDAR systems (near the telescopes) and with
a laser facility near the center of the array.\cite{roberts,malek}
The Cherenkov light contribution of the shower is calculated
according to Ref.~\cite{nerling}.
An energy deposit profile is
reconstructed for each event. 
A Gaisser-Hillas function \cite{gh} is fitted to the
profile to obtain the depth of shower maximum, and
the calorimetric shower energy is obtained by integration.
A 1\% correction for missing energy assuming photon primaries \cite{pierog}
is applied.

The following quality cuts are applied for event selection:
(i) quality of hybrid geometry: distance of closest approach of the
reconstructed shower axis to the array tank with the largest signal
$<$1.5~km, and difference between the reconstructed shower front
arrival time at this tank and the measured tank time $<$300~ns,
(ii) number of phototubes in the fluorescence detector
                 triggered by shower $\ge$6,
(iii) quality of Gaisser-Hillas (GH) profile fit:
    $\chi^2$(GH) per degree of freedom $<$6, and
    $\chi^2$(GH)/$\chi^2$(line)$<$0.9, where $\chi^2$(line) refers
    to a straight line fit,
(iv) minimum viewing angle of shower direction
                 towards the telescope $>$15$^\circ$,
(v) primary energy E$>$10$^{19}$~eV,
(vi) $X_{\rm max}$ observed in the field of view,
(vii) cloud monitors confirm no disturbance of
                 event observation by clouds.

\begin{figure}
\epsfig{figure=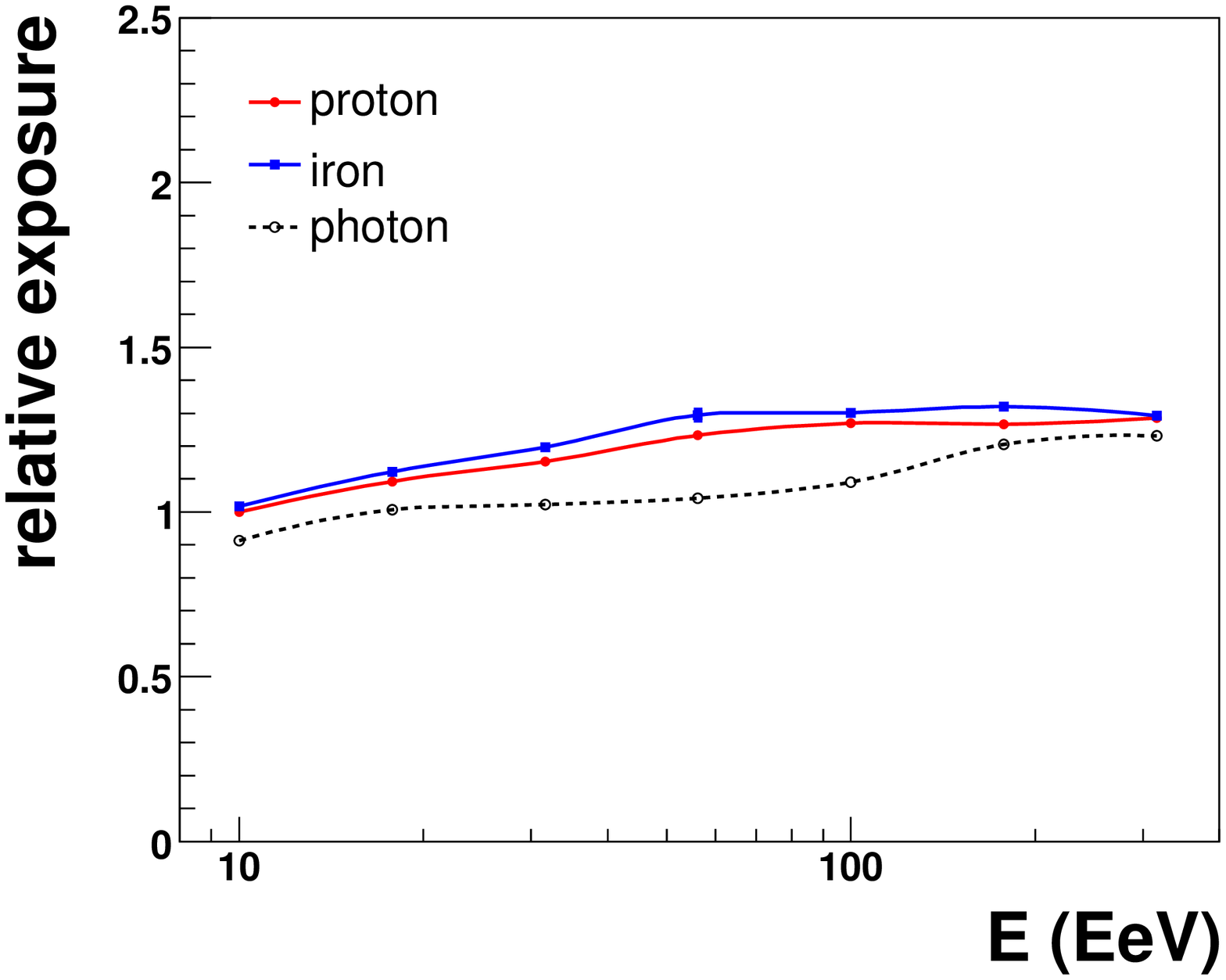,height=1.6in}
\epsfig{figure=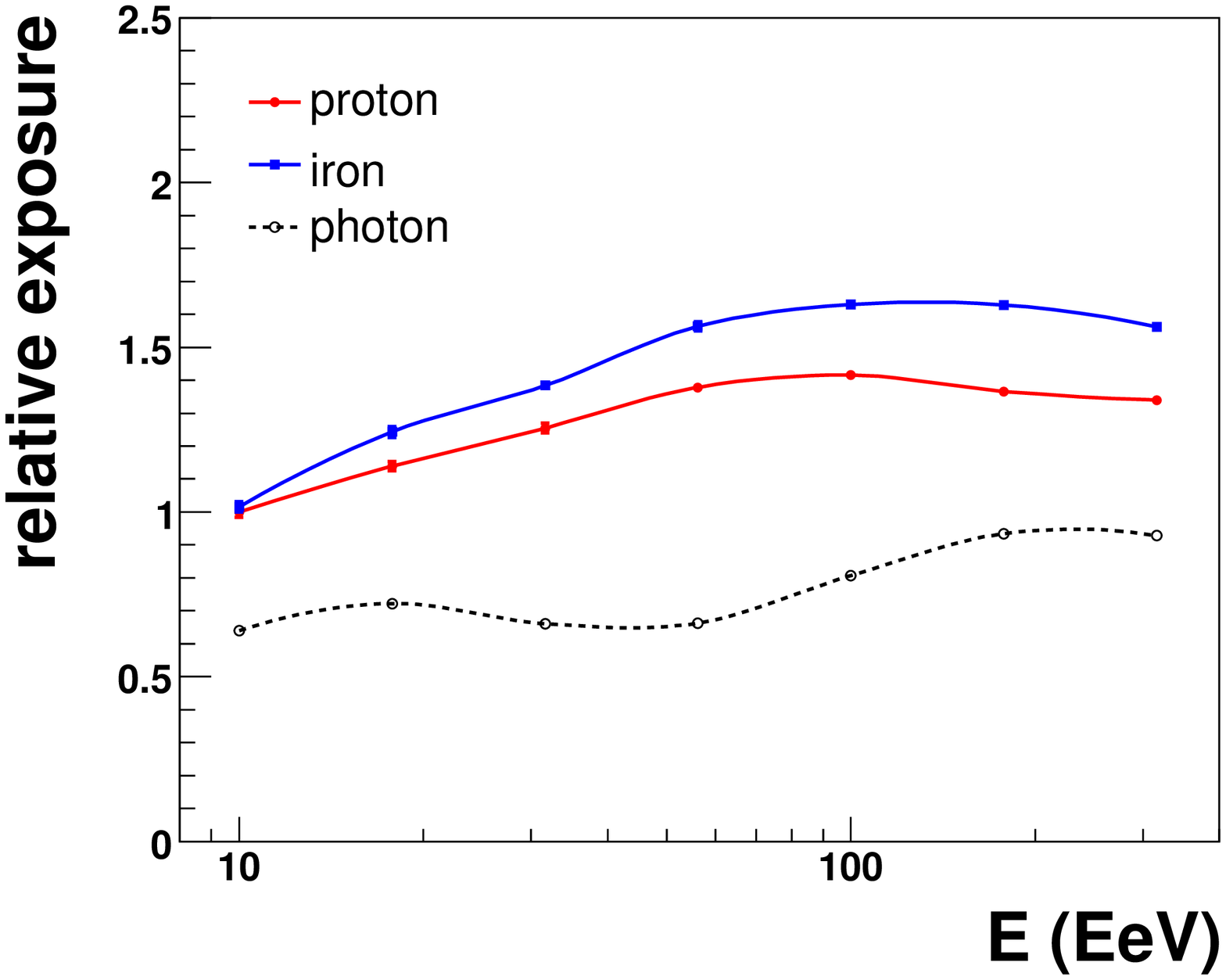,height=1.6in}
\epsfig{figure=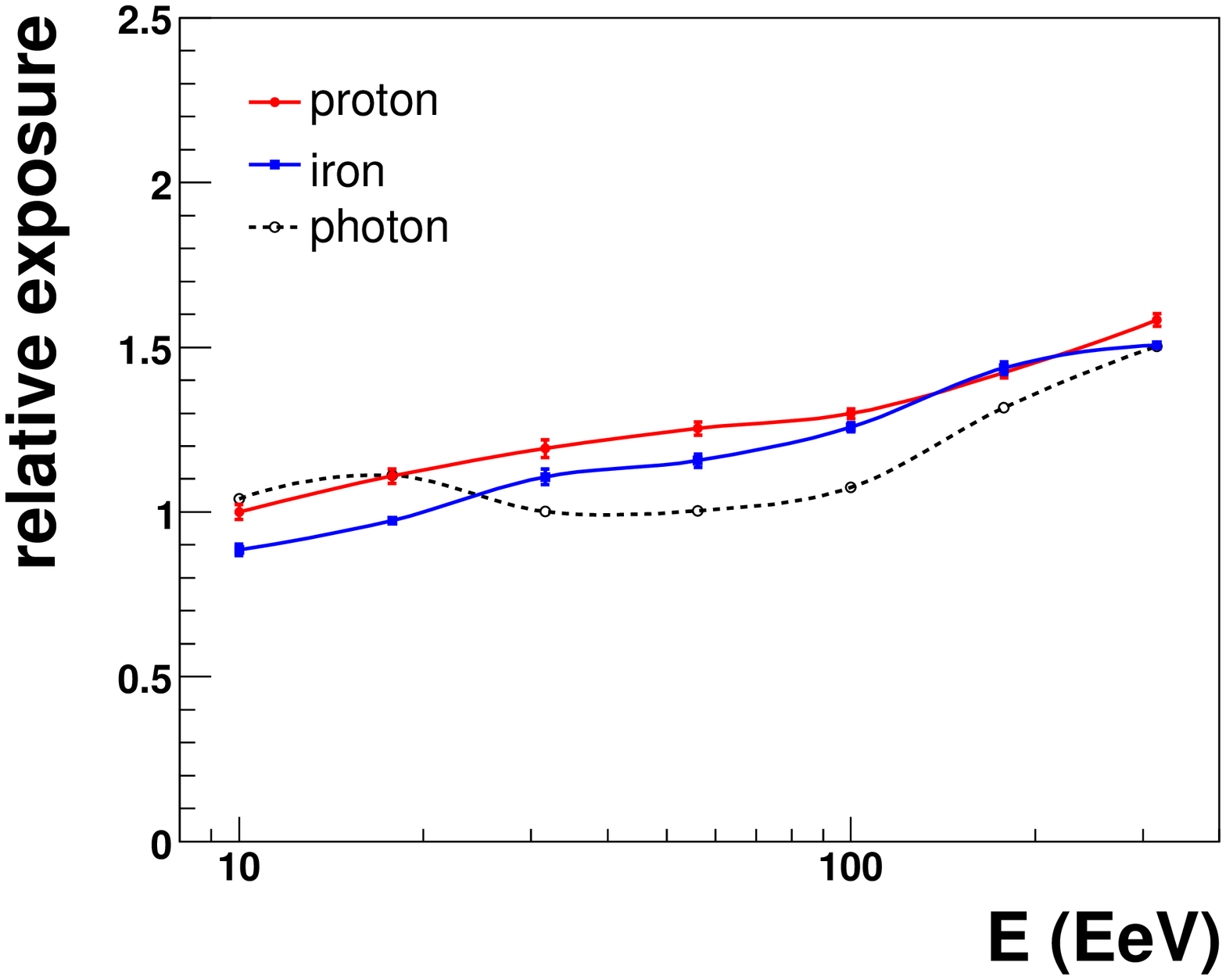,height=1.6in}
\caption{
Relative exposures for photon, proton, and iron primaries
as a function of energy after trigger (left), after quality cuts (middle)
and after fiducial volume cuts are applied (right) to reduce the bias
against photons.
A reference value of one is adopted for proton at 10~EeV.
\label{fig-acc}}
\end{figure}

Care must be taken about a possible bias of the detector acceptance
against photon primaries.
In Figure~\ref{fig-acc} we show the acceptance for photons and
nuclear primaries at different steps of the analysis, computed
using shower simulations 
with the CONEX code \cite{conex} which reproduces well the CORSIKA
predictions for shower profiles. 
Light emission and propagation through the atmosphere and
the detector response were simulated according to Ref. \cite{prado}.
As can be seen from the Figure, the acceptances are comparable for
all types of primaries after trigger (left plot).
However, after profile quality cuts
(middle plot) the detection efficiency for photons is smaller by
a factor $\sim$2 than for nuclear primaries, because primary photons reach
shower maximum at such large depths (of about 1000~g~cm$^{-2}$,
see Figure~\ref{fig-xmaxvse}) that
for a large fraction of showers the maximum is outside the field of
view of the telescopes.
To reduce the corresponding bias against photons, near-vertical events
are excluded in the current analysis.
Since the average depth of shower maximum increases with
photon energy before the onset of preshower,
a mild dependence of the minimum zenith angle with energy
is chosen. In addition, an energy-dependent distance cut
is applied to the data as the telescopes do not observe shower
portions near the horizon:
(i) zenith angle
                 ~$>$35$^\circ~+~g_1(E)$, with
                 $g_1(E)= 10(\lg E/$eV$-19.0)^\circ$
                 for $\lg E/$eV$\le19.7$ and
                 $g_1(E)=7^\circ$ for $\lg E/$eV$>$19.7,
(ii) maximum distance of telescope to shower impact point
                 ~$<$24~km~+~$g_2(E)$, with
                 $g_2(E)= 12(\lg E/$eV$-19.0)$~km.

The acceptances after application of the fiducial volume cuts are shown in
Figure~\ref{fig-acc} (right plot).
The differences between photons and nuclear primaries are now
significantly reduced, with the acceptances being comparable
at energies 10--20~EeV.
With increasing energy, the acceptance for nuclear primaries shows
a modest growth, while the photon acceptance is quite flat in
the investigated energy range. 
Comparing photons to nuclear primaries, the minimum ratio of 
acceptances is $\epsilon_{\rm min} \simeq 0.80$ at energies 50--60~EeV.
At even higher energies, the preshower
effect becomes increasingly important, and acceptances for
photons and nuclear primaries become more similar.
To obtain an experimental limit to the photon fraction
without relying on assumptions on energy spectra of different primaries,
a correction to the photon limit is applied
by conservatively adopting the minimum ratio of acceptances
$\epsilon_{\rm min}$ (a detailed derivation of the approach
is given in Ref. \cite{paper}).

Applying the cuts to the data, 29 events with energies
greater than 10~EeV satisfy the selection criteria.
Due to the steep cosmic-ray
spectrum, most events in the sample (23 out of 29) are below 20 EeV.
Figure~\ref{fig-evprof} (left plot) shows the longitudinal profile of an event
reconstructed with 16~EeV and $X_{\rm max} = 780$~g~cm$^{-2}$.
The $X_{\rm max}$ values of the selected events are displayed
in Figure~\ref{fig-xmaxdata}.
A Table summarising the main shower characteristics for all events
can be found in Ref. \cite{paper}.

The uncertainty $\Delta X_{\rm max}$ of the reconstructed depth of
shower maximum is composed of several contributions, some of which
may vary from event to event.
In this work, we conservatively adopt overall estimates for the current
statistical and systematic uncertainties which
are applied to all selected events.
These uncertainties are expected to decrease significantly in the future.
However, even when adopting conservative estimates,
the present analysis is not limited by the measurement
uncertainties but by event statistics.
This is due to the fact that shower fluctuations for photons are
considerably larger than the measurement uncertainties.

Main contributions to $\Delta X_{\rm max}$ of 10$-$20~g~cm$^{-2}$ each
are the uncertainties in the profile fit, in shower geometry, and in
atmospheric conditions (see Table in Ref. \cite{paper}).
The current systematic uncertainty of 25\% in energy 
reconstruction~\cite{bellido}
translates to a $\sim$~13~g~cm$^{-2}$ systematic uncertainty
in the $X_{\rm max}^{\gamma}$ values predicted from photon simulations.
The uncertainty from extrapolating the photo-nuclear cross-section
to high energies is estimated to $\sim$~10~g~cm$^{-2}$. \cite{strikman,rissec2cr}
Contrary to the case of nuclear primaries, uncertainties from
modelling  high-energy hadron interactions are much less important
in primary photon showers (uncertainty of $\sim$~5~g~cm$^{-2}$): this is
a great advantage of this type of analysis where data are compared to calculations
of primary photon showers only.
Adding in quadrature the individual contributions
gives a statistical uncertainty
$\Delta X_{\rm max}^{\rm stat}\simeq$ 28~g~cm$^{-2}$
and a systematic uncertainty
$\Delta X_{\rm max}^{\rm syst}\simeq$ 23~g~cm$^{-2}$.

For each selected event, 100 showers were simulated as photon primaries.
Since photon shower features can depend in a non-trivial way
on arrival direction and energy, the specific event conditions
were adopted for each event.

\section{Results}

\begin{figure}
\epsfig{figure=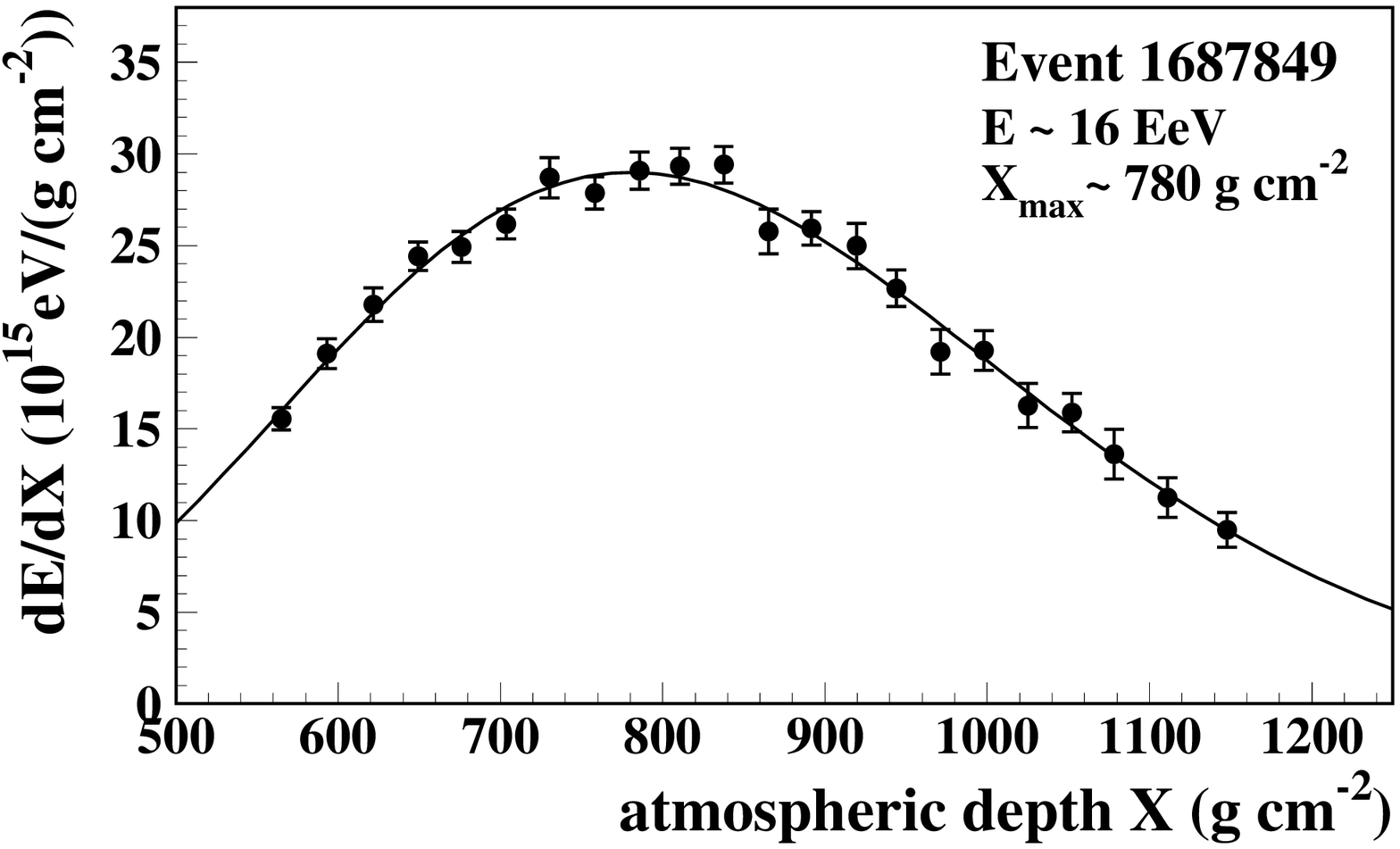,height=1.9in}
\epsfig{figure=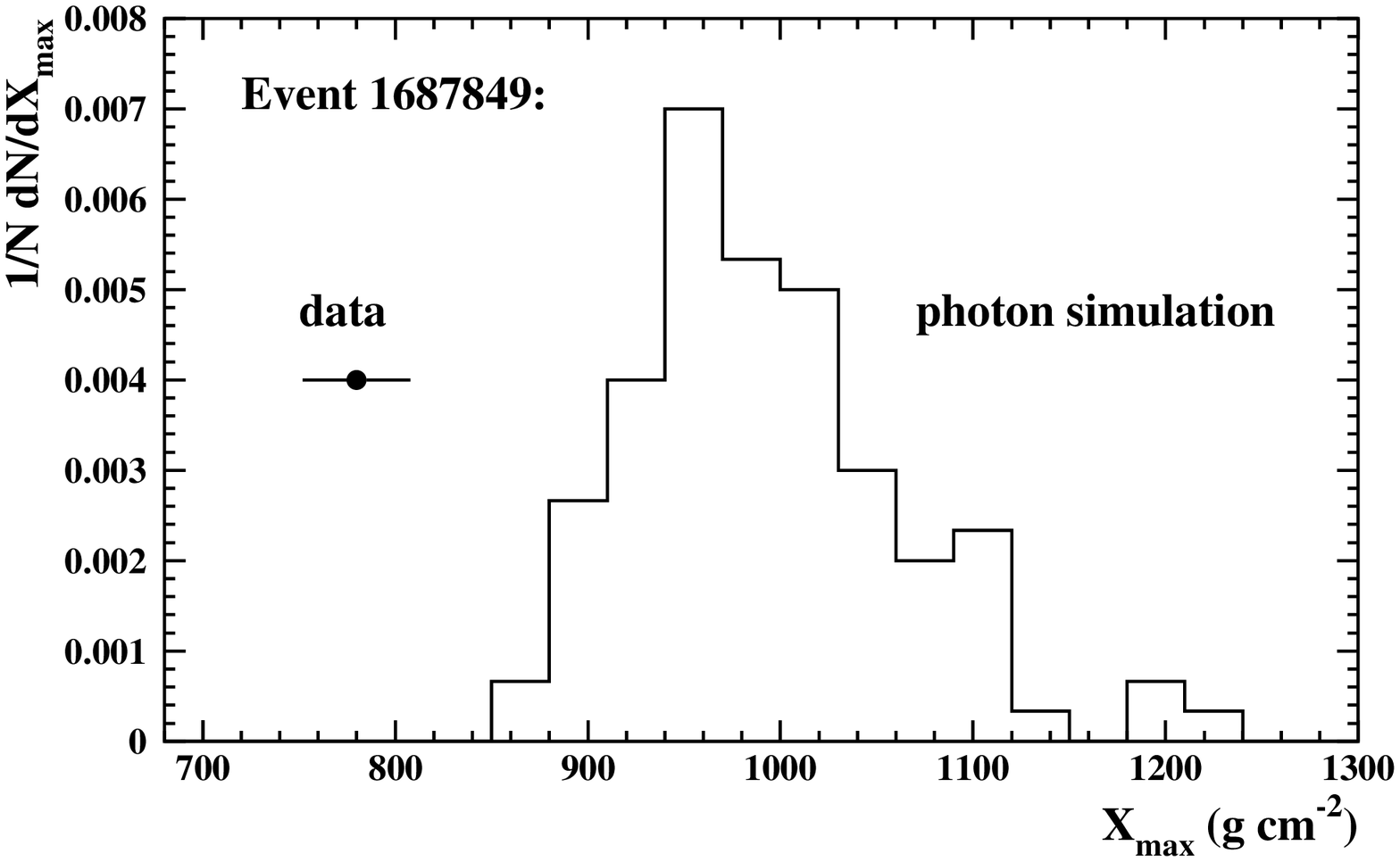,height=1.9in}
\caption{
{\it Left:} example of a reconstructed longitudinal energy deposit profile
(points) and the fit by a Gaisser-Hillas function (line).
{\it Right:}
$X_{\rm max}$ measured in the shower shown on the left-hand side
(point with error bar)
compared to the $X_{\rm max}^\gamma$ distribution
expected for photon showers (solid line).
\label{fig-evprof}}
\end{figure}

In Figure~\ref{fig-evprof} (right plot) the predictions for $X_{\rm max}^\gamma$
for a photon primary are compared with the measurement of 
$X_{\rm max} = 780$~g~cm$^{-2}$
for the event shown in the left plot of Figure~\ref{fig-evprof}.
With $\langle X_{\rm max}^\gamma \rangle \simeq 1000$~g~cm$^{-2}$,
photon showers are 
on average expected to reach maximum at depths considerably greater than
that observed for real events.
Shower-to-shower fluctuations are large due to the LPM effect.
For this event, the expectation for a primary photon differs by
$\Delta_\gamma \simeq$ +2.9 standard deviations from the data,
where $\Delta_\gamma$ is calculated from
\begin{equation}
\label{eq1}
\Delta_\gamma = \frac{ <X_{\rm max}^\gamma> - X_{\rm max}}
{ \sqrt{ (\Delta X_{\rm max}^\gamma)^2 +
  (\Delta X_{\rm max}^{\rm stat})^2 } }~~.
\end{equation}

\begin{figure}
\epsfig{figure=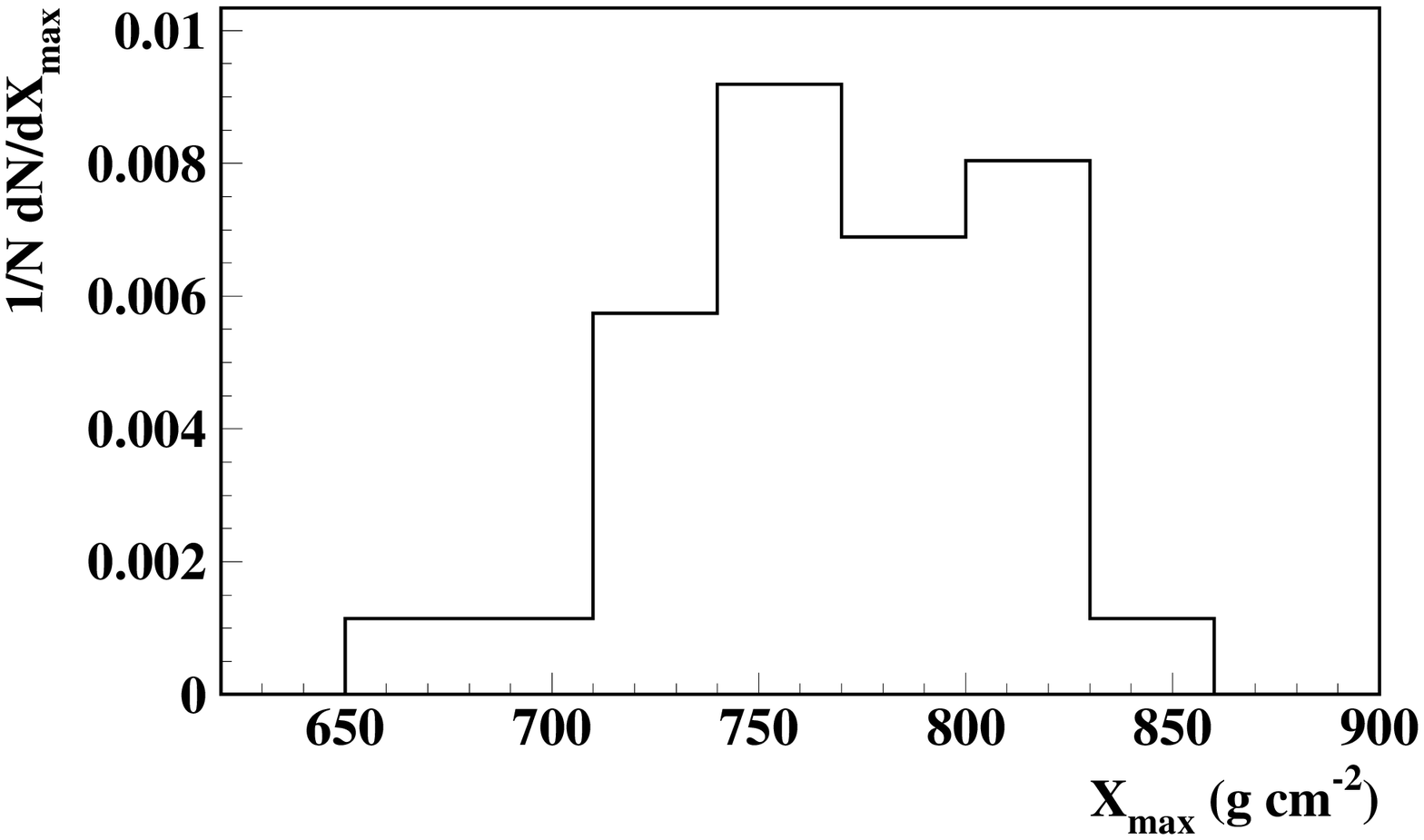,height=1.9in}
\epsfig{figure=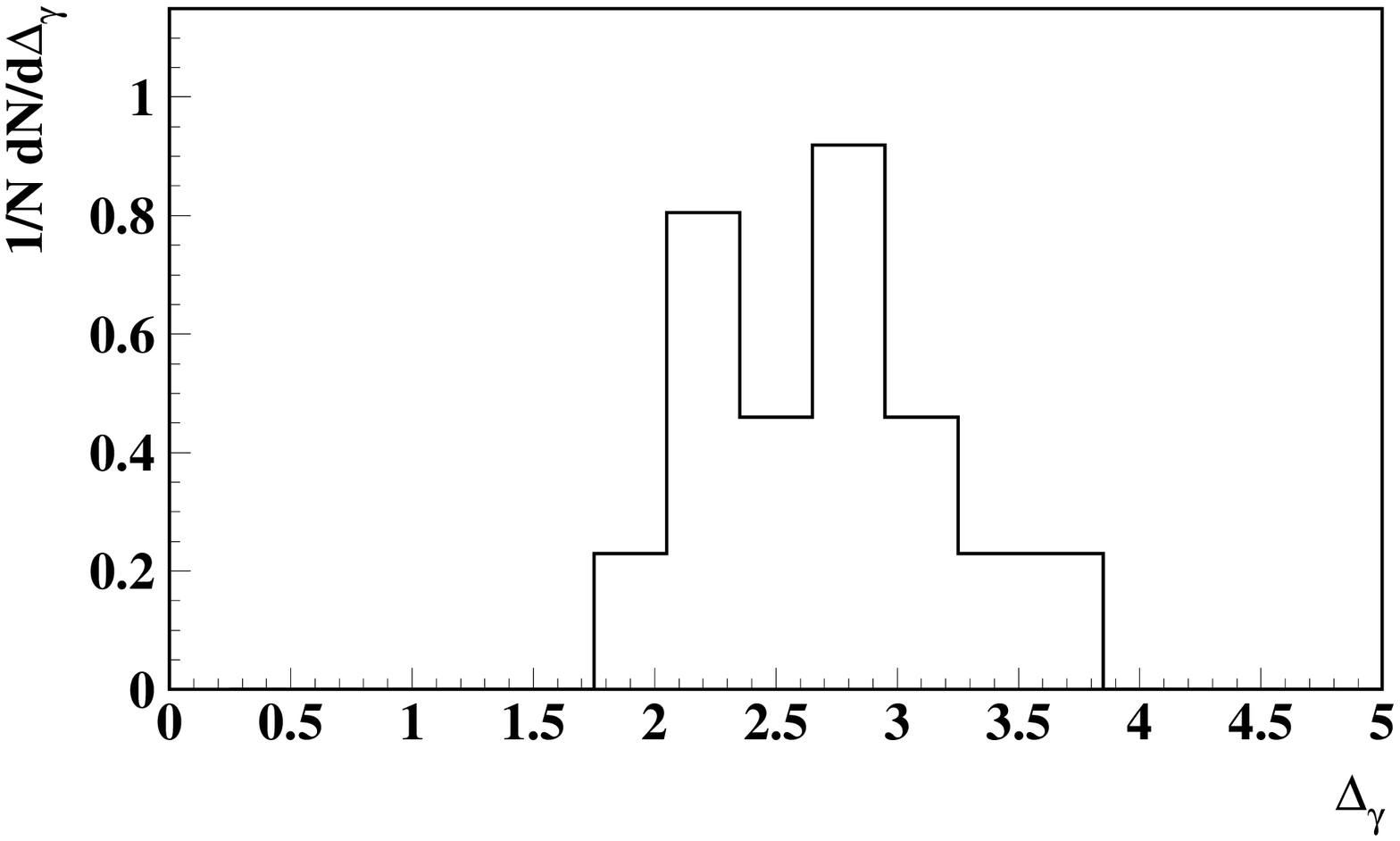,height=1.9in}
\caption{
{\it Left:} Distribution of $X_{\rm max}$ values of the 29 selected events.
{\it Right:} distribution of differences $\Delta_\gamma$ in standard deviations
between primary photon prediction and data for the 29 selected events.
\label{fig-xmaxdata}}
\end{figure}

For all events, the observed $X_{\rm max}$ is well below the average
value expected for photons. The distribution of observed $X_{\rm max}$ values
is shown in Figure~\ref{fig-xmaxdata} (left plot).
The differences $\Delta_\gamma$ between photon prediction and data
range from +2.0 to +3.8 standard deviations,
see Figure~\ref{fig-xmaxdata} (right plot).
It is extremely unlikely that all 29 events were initiated by photons
(probability $\ll$10$^{-10}$),
so an upper limit to the fraction of cosmic-ray photons above 10~EeV
can be reliably set.
Due to the limited event statistics, the upper limit cannot be
smaller than a certain value.
For 29 events and $\epsilon_{\rm min} \simeq 0.80$, the minimum
possible value for an upper limit
to be set at a 95\% confidence level is $\sim$~12\%.
The theoretical limit is reached only if
a photon origin is basically excluded for all events.

The calculation of the upper limit is based on the statistical
method introduced in Ref. \cite{risse05} which is tailor-made for
relatively small event samples.
For each event, trial values $\chi^2 = \Delta_\gamma^2$ are calculated
with $\Delta_\gamma$ according to Eq.~(\ref{eq1}).
We distinguish between statistical and systematic uncertainties for
the depths of shower maximum.
The method in Ref. \cite{risse05} is extended to allow for a correlated
shift of the observed $X_{\rm max}$ values for all selected events,
where the shifted value is drawn at random from a Gaussian distribution
with a width $\Delta X_{\rm max}^{\rm syst}$ = 23~g~cm$^{-2}$.
For the shifted data, new $\chi^2$ values are calculated from
Eq.~(\ref{eq1}).
Many such ``shifted'' event sets are generated from the data and 
compared to artificial data sets using photon simulations.
The chance probability $p(f_\gamma)$ is calculated to obtain artificial
data sets with $\chi^2$ values larger than observed as a function of the
hypothetical primary photon fraction $f_\gamma$.
Possible non-Gaussian shower fluctuations are accounted for in the
method, as the probability is constructed by a Monte Carlo technique.
The upper limit $f_\gamma^{\rm ul}$, at a confidence level $\alpha$, is
then obtained from 
$p (f_\gamma \ge \epsilon_{\rm min} f_\gamma^{\rm ul}) \le 1-\alpha$,
where the factor $\epsilon_{\rm min} = 0.80$ accounts for the different
detector acceptance for photon and nuclear primaries. 

\begin{figure}
\begin{center}
\epsfig{figure=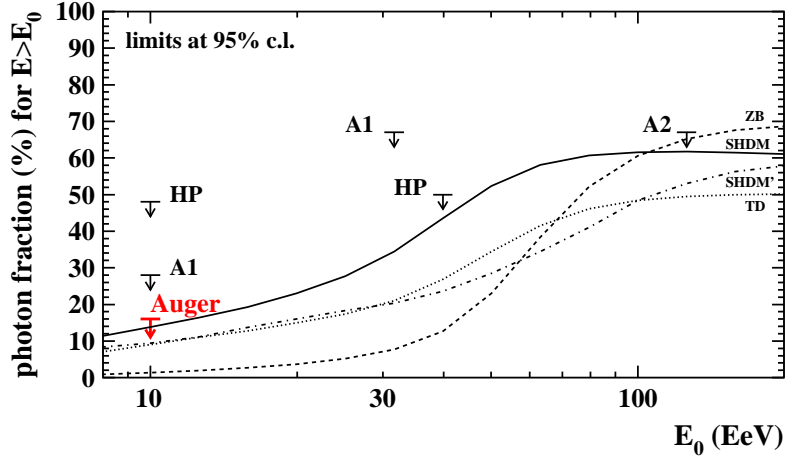,height=2.5in}
\caption{
Upper limits (95\% c.l.) to the cosmic-ray photon fraction
derived in the present analysis (Auger) and obtained
previously from AGASA (A1),$^{38}$ (A2),$^{39}$
and Haverah Park (HP) $^{40}$ data, compared to expectations
for non-acceleration models
(ZB, SHDM, TD from Ref. $^{10}$, SHDM' from Ref. $^{11}$).
\label{fig-uplim}}
\end{center}
\end{figure}

For the Auger data sample, an upper limit to the
photon fraction of 16\% at a confidence level of 95\% is derived.
In Figure~\ref{fig-uplim}, this upper limit is plotted together with
previous experimental limits and some illustrative estimates for
non-acceleration models. We have shown two different expectations for SHDM decay 
\cite{models,ellis} to illustrate the sensitivity to assumptions made about the
decay mode and the fragmentation, as well as the normalisation of the spectrum. 
The derived limit is the first one based on observing the depth
of shower maximum with the fluorescence technique.
The result confirms and improves previous limits above 10~EeV
that were derived from surface arrays.
It is worth mentioning that this improved limit is achieved 
with only 29 events above 10~EeV,
as compared to about 50 events in the Haverah Park analysis and about 120 events
in the AGASA analysis.

\section{Outlook}

The current analysis is limited mainly by the small number of events.
The number of hybrid events will considerably increase over the
next years, and much lower primary photon fractions can be tested.
With a data increase of about an order of magnitude compared to the current
analysis, as is expected to be reached in 2008/2009, photon fractions of
$\sim$~5\% can be tested.
Moreover, the larger statistics will allow us to increase the
threshold energy above 10~EeV where even larger photon fractions
are predicted by most models.
A comparable number of events as for the present analysis would be reached
then above 30--35~EeV.
An upper limit similar to the current one but at this higher energy 
would severely constrain non-acceleration models.

\begin{figure}
\begin{center}
\epsfig{figure=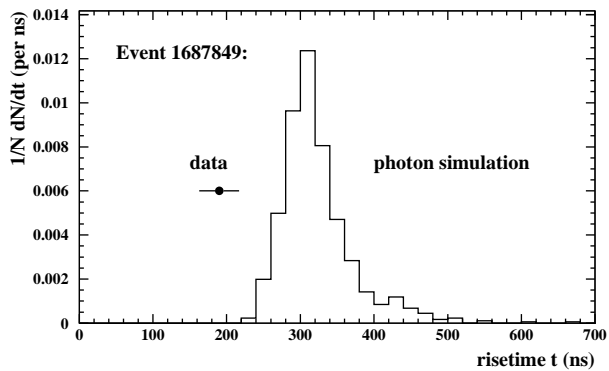,height=2.0in}
\caption{
Example of risetime measured in an individual shower,
same as in Figure~\ref{fig-evprof}
(point with error bar) compared to the risetime distribution
expected for photon showers (solid line).
\label{fig-evrise}}
\end{center}
\end{figure}

In this work, data from the surface array are used
only to achieve a high precision
of reconstructed shower geometry in hybrid events. A single tank was
sufficient for this. However, observables registered by the surface
array are also sensitive to the primary particle type and can be
exploited for studies of primary photon showers.\cite{bertou00}
An example for another observable is given by the {\it risetime}
of the shower signal in the detectors, one measure of the time
spread of particles in the shower disc.
For each triggered tank, we define a risetime
as the time for the integrated signal to go from 10\% to 50\% of its
total value.
By interpolation between risetimes recorded by the tanks at different
distances to the shower core, the risetime at 1000~m core distance
is extracted after correcting
for azimuthal asymmetries in the shower front.
For the specific event shown in Figure~\ref{fig-evprof},
the measured risetime 
is compared to the simulated distribution in Figure~\ref{fig-evrise}. 
The observed risetime does not agree well with the predictions
for primary photons.
In future photon analyses, the independent information
on the primary particle from the Auger ground array and fluorescence
telescope data can be used to cross-check each other.
Combining the different shower observables will further
improve the discrimination power to photons.

If only surface detector data are used, event statistics are increased
by about an order of magnitude.  However, care must be taken about a
possible bias against photons in an array-only
analysis because of the different detector acceptance for photon and nuclear
primaries. Also, compared to the near-calorimetric energy determination
in the fluorescence technique, the energy estimated from array data
shows a stronger dependence on the primary type and is more strongly
affected by shower fluctuations.
These issues are under current investigation.

It is planned to complement the southern site of the Auger Observatory
by a northern one in Colorado. This can substantially improve the sensitivity
to UHE photons. It is then important to note that UHE photons are expected
to be produced during propagation even if the sources emit nucleons only,
resulting in a ``guaranteed'' flux of UHE photons.\cite{models}
Detection of these photons would be a major step forward for
investigating sources and interactions at UHE.

Comparing the northern and southern Auger sites, differences in the expected
preshower features exist.\cite{homola_ns}
Due to the stronger (factor $\sim$2) magnetic field, the preshower
process starts at smaller energies at the northern site.
The ``sky pattern'' for preshowering (probability
of geomagnetic pair production for a given energy as a function of the
arrival direction) is shifted according to the different pointing of the local
magnetic field lines. Interestingly, values of $\sim$100\% preshower
probability are reached for the full sky at higher energies in Colorado
(despite the stronger magnetic field on ground), which is connected
to the field lines being less curved with distance from the ground.
These differences may be exploited to search for UHE photons or,
in particular, to obtain independent proof of a possible photon signal
detected at one site.

\section*{Acknowledgments}
It is a pleasure to thank the organizers of the {\it Rencontres du Vietnam}
for their invitation and for a very pleasant and informative conference
atmosphere.
The author is grateful to K.-H.~Kampert for helpful comments on the
manuscript.
Partial support from the Helmholtz VIHKOS Institute and from the
German Ministry for Education and Research is kindly acknowledged.

\end{document}